\documentclass{IEEEtran}

\usepackage{amsmath,amssymb,amsfonts,bbm}

\DeclareMathAlphabet{\mathsfbf}{OT1}{cmss}{sbc}{n}
\usepackage{cite}

\usepackage[tight,footnotesize]{subfigure}

\usepackage{times,url,comment,cite}
\usepackage{amsmath,amssymb,amsfonts,bbm,bbold}
\usepackage{graphicx,slashbox}
\usepackage{color}
\usepackage{subfigure}
\usepackage{comment}
\usepackage{algorithm2e}

\begin{document}

\title{Secure Neighbor Position Discovery in VANETs}

\author{\IEEEauthorblockN{M. Fiore, C. Casetti, C.-F. Chiasserini, and}
\IEEEauthorblockN{P. Papadimitratos}
}

\maketitle

\begin{abstract}

Many significant functionalities of vehicular ad hoc networks (VANETs) require
that nodes have knowledge of the positions of other vehicles, and notably of
those within communication range. However, adversarial nodes could provide false
position information or disrupt the acquisition of such information. Thus, in
VANETs, the discovery of neighbor positions should be performed in a secure
manner. In spite of a multitude of security protocols in the literature,  there
is no secure discovery protocol for neighbors positions.  We address this
problem in our paper: we design a distributed protocol that relies solely on
information exchange among one-hop neighbors, we analyze its security properties
in presence of one or multiple (independent or colluding) adversaries, and we
evaluate its performance in a VANET environment using realistic mobility traces.
We show that our protocol can be highly effective in detecting falsified
position information, while maintaining a low rate of false positive detections.


\end{abstract}

\section{Introduction}

%

VANETs are envisioned to enable a range of applications, spanning from enhanced
transportation safety and efficiency to mobile infotainment,
while security and privacy enhancing technologies have been broadly accepted as prerequisites
for the deployment of such systems. A number of on-going efforts have yielded a
multitude of proposed schemes, including coordinated efforts such as those of
the IEEE 1609 working group, the Car-to-Car Communication Consortium, the
CAMP/VSC-2 project, and the SeVeCom project, which produced a full-fledged
security architecture for vehicle-to-vehicle and vehicle-to-infrastructure
communications.

Many aspects of security and privacy have already been  addressed (e.g.,
in~\cite{Shen1,Shen2,Shen3}) but no solution has been yet proposed for the secure
discovery of the position of other nodes, in particular those within direct
communication range. This is an important problem because vehicular nodes are
location-aware, and location information is embedded in many VANET messages to
support various applications; transportation safety and geographical forwarding
(or GeoCast) are characteristic examples, while traffic monitoring and
management, as well as access to location-based services are also closely
related. In all such cases, nodes are required to reliably identify neighboring
nodes and determine their positions. Nonetheless, adversarial or faulty nodes
can falsify or alter such information, resulting in the disruption of system
operations.


Secure discovery of the positions of neighbors cannot be achieved by any of the
solutions in the literature. Secure localization techniques, which allow a
reliable determination of own location, are a building block but not the
solution to the  problem at hand. Simply put, the reason is that an adversary
could advertise a false position in any discovery protocol. The presence of
trusted nodes would make the problem easier to solve: road-side infrastructure
or trustworthy specialized vehicles could help to securely localize other
vehicles. In such case, techniques in the literature, designed for mobile ad-hoc
networks, could be employed. However, this approach has severe limitations when
applied to vehicular environments: the presence of road-side infrastructure is
envisioned to be rather sparse and the presence of trustworthy nodes cannot be
guaranteed at all times, whereas position discovery is needed at any time and
location among any two or more vehicles.

To address this problem, we propose our Secure Neighbor Position Discovery
(SNPD) protocol, which enables any node (i) to discover the position of its 
neighbors on-demand and in real-time; and (ii) to detect and discard faulty positions
and, thus, ignore their originators. SNPD therefore allows any vehicular node to 
autonomously obtain a set of verified neighbor positions, leveraging the 
contributions of its peers to weed out wrong-doers, without any prior 
assumption about their trustworthiness.


In the rest of the paper, we first discuss related work and introduce
the system and adversary model we adopt, then we describe our SNPD protocol in
detail. A security analysis of SNPD follows, along with a performance
evaluation based on realistic vehicular mobility traces.

\section{Related Work\label{sec:related}}

Secure neighbor position discovery for vehicular environments is,
to the best of our knowledge, an open problem. Nevertheless, it relates to a
number of other problems that have instead been addressed before,
as discussed next. We emphasize that
our SNPD protocol is compatible with state-of-the-art security architectures
for vehicular networks, including those proposed by IEEE 1609.2~\cite{Ieee05:_IEEE_P1609.2} 
and SeVeCom~\cite{Papadimitratos08:_secure_veh_comm_arch}.

\textbf{Securing own location and time information} is orthogonal to our problem,
as adversaries can acquire their own locations in a reliable manner, but then
advertise false positions to their neighbors. Own positioning and time synchronization
is thus a building block for SNPD, as it is for secure vehicular networking.
In vehicular environments, self-localization is mainly
achieved through Global Navigation Satellite Systems, e.g., GPS, whose security
can be provided by cryptographic and non-cryptographic defense 
mechanisms~\cite{milcom2008}; alternatively, other terrestrial special-purpose
infrastructure (beacons) could be used~\cite{poovendran07}, along with techniques
to deal with non-honest beacons~\cite{zhong:theoryrobust}. In the rest of this
paper, we assume that devices can determine securely their own position and time
reference.

\textbf{Secure neighbor discovery (SND)}, that is, the discovery of directly
reachable nodes (communicating neighbors) or nodes within a distance (physical
neighbors)~\cite{snd-mag}, is only a step towards the solution we are after. To
put it simply, an adversarial node could be securely discovered as neighbor and
be indeed a neighbor (within some SND range), but it could still cheat about its
position within the same range. SND is a subset of the SNPD  problem, since it
lets a node assess whether another node is an actual neighbor but it does not
verify the location it claims to be at. Nonetheless, properties of SND
protocols  with proven secure solutions~\cite{asiaccs08,fmse-ccs}, are useful
in our context: as an example, signal Time of Flight-based and other distance
measurements between two nodes can prevent relay attacks (i.e., malicious nodes
relaying, stealthily and verbatim, messages of other correct nodes).


\textbf{Neighbor position verification} was investigated in the context of 
ad-hoc networks, with solutions relying on dedicated mobile or hidden base 
stations~\cite{capkun08}, or on the availability of a number of  trustworthy
devices~\cite{capkun:secpos}. Our SNPD protocol, instead, is a fully 
distributed solution that does not require the presence of any particular
infrastructure or a-priori trusted neighbors. Also, unlike previous works, our
solution targets highly mobile  environments and it only assumes RF
communication; indeed, non-RF communication, e.g., infra-red or ultra-sound, is
unfeasible in VANETs, where non-line-of-sight conditions are frequent and
car-to-car distances often are in the order of tens or hundreds of meters.


\section{System and adversary model}
\label{sec:model}

We consider a vehicular network whose nodes communicate over a high-bit-rate
data link through an RF interface.
We assume that each node knows its own location with
some maximum error $\epsilon_p$, and that it shares a common time reference
with the other nodes in the network: both requirements can be met by equipping
vehicles with GPS receivers, already a major trend in today's car 
manufacturing\footnote{With the help of GPS, user synchronization, 
fine time granularity and a relatively precise
location information is available. Currently, small-footprint and
low-cost GPS receivers are commercially available, which achieve low
synchronization error and low localization error.}.
Also, nodes can perform Time of Flight (ToF)-based RF ranging using one message
transmission, with a maximum error equal to $\epsilon_r$: as discussed
in~\cite{capkun:secpos,techrep}, this is a reasonable assumption, although it requires
modifications to the current off-the-shelf radio interfaces; $\epsilon_p$
and $\epsilon_r$ are assumed to be equal for all nodes.

Each node has a unique identity, and carries cryptographic keys that allow it
to authenticate messages from other nodes in the network.
Although there are various ways to enable authentication, here
we only require that message authentication is done locally
and we assume that each node $X$ holds its own pair of private and public keys,
$k_X$ and $K_X$, respectively, as well as a set of one-time use keys
\{$k'_X, K'_X$\}. $X$ can encrypt and decrypt data with
its key(s) and the public keys of other nodes; also, it can
produce digital signatures with its private key. 
We assume that the binding between $X$ and $K_X$ can be validated by any node,
as in state-of-the-art vehicular communication architectures.

Nodes either comply with the SNPD protocol (\emph{correct}) or they deviate from
it (\emph{faulty} or \emph{adversarial}).  Adversarial nodes can advertise
arbitrarily erroneous positions in messages they inject, to mislead other nodes
about their position.
Adversaries are \emph{external} or \emph{internal}, depending on whether they
lack or possess the cryptographic keys and credentials of system nodes,
respectively. External adversaries can only relay or replay messages without
changes, or jam the communication. Internal adversaries are more powerful in
that they can fully participate in the protocol execution, forging arbitrary
messages with faked own positions.  Recall though that each adversary can inject
messages only according to the cryptographic keys it possesses; it cannot forge
messages on behalf of other nodes whose keys it does not have.
Another classification of adversaries that is of interest to us is between
\emph{independent} and \emph{colluding} 
adversaries: the former act without knowledge of 
other adversaries in the neighborhood, while the latter, by far the most 
dangerous, coordinate their actions by exchanging information.

In this work, we focus primarily on internal adversaries with standard 
equipment (e.g., omnidirectional antennas, standard--compliant wireless
cards, etc.). We distinguish them into (i) {\em knowledgeable}, i.e., adversaries
that at any point in time know the exact positions of all their communication
neighbors, and (ii) {\em unknowledgeable}, otherwise. In
Section~\ref{sec:analysis}, we will outline the threats which  can be posed
by both independent and colluding adversaries, and discuss possible additional 
threats carried out by adversaries using non-standard equipment (e.g.,
directional antennas).



\section{Secure neighbor position discovery protocol}
\label{sec:protocol}

The SNPD protocol we propose allows any node in the network to discover and
verify the position of its communication neighbors participating in the
protocol message exchange. SNPD can be initiated in a reactive manner by any
node, which we refer to as the {\em verifier}. Our solution is based on a
\emph{best-effort, cooperative approach} that leverages information collected by
neighboring nodes thanks to the broadcast nature of the wireless medium. With
such information, the verifier can compute, via ToF-based ranging, distances
between pairs of neighbors, and then perform a sequence of tests that allow it
to classify its communication neighbors as:

\begin{itemize}

\item {\em Verified}, i.e., nodes the verifier deems to be at the claimed
position;

\item {\em Faulty}, i.e., nodes the verifier deems to have announced an
incorrect position;

\item {\em Unverifiable}, i.e., nodes the verifier cannot prove to be either
correct or faulty; due to insufficient information on these nodes or
inconclusive test outcome. 

\end{itemize}

The objective of our SNPD protocol is to be robust to adversarial nodes, i.e.,
to correctly identify and reject false positions and ignore their originators.
In other words,
it is necessary to minimize false negative and false positive outcomes, i.e.,
adversaries with positions deemed verified and correct nodes with positions
deemed faulty, as well as the number of unverifiable nodes.

We stress that the SNPD protocol only verifies the position of those neighbors
with which the message exchange takes place successfully. It therefore
disregards nodes for which the protocol exchange prematurely ends, e.g., due to
message loss or communication neighbors that refuse to take part in the
protocol. SNPD assumes that the nodes position does not vary significantly
during one protocol execution, which is realistic if we consider that a complete
message exchange takes no more than a few hundreds of milliseconds. 
Also, SNPD does not aim at building a consistent map of verified nodes, as 
every verifier autonomously tags its neighbors as verified, faulty or
unverifiable.

Next, we
detail the message  exchange between the verifier and its communication
neighbors, followed by a description  of the security tests run by the verifier.
Table~\ref{tab:notation} summarizes the notations used throughout the protocol
description.


\subsection{Message exchange}
We denote by $t_X$ the time at which a node $X$ starts a broadcast
transmission and by $t_{XY}$ the time at which a node $Y$ starts
receiving that same transmission; $p_X$ is the current position of $X$,
and $\mathbb{N}_X$ is the current set of its communication neighbors. 
Consider a verifier $S$ that initiates the SNPD protocol.
The message exchange procedure is outlined in Algorithm~\ref{alg:exchange_init}
for  $S$, and in Algorithm~\ref{alg:exchange_neigh} for 
any of $S$'s communication neighbors.

The verifier starts the protocol by broadcasting a {\sc poll} whose transmission
time $t_S$ is stored locally (Alg.~\ref{alg:exchange_init}, lines 2-3). Such
message is anonymous, since (i) it does not contain the verifier's identity,
(ii) it is transmitted employing a fresh MAC address, and (iii) it contains a
public key $K_{S}'$ from a one-time use private/public key pair $k_{S}',K_{S}'$,
taken from a pool of anonymous keys which do not allow neighbors to map them
onto a specific node. Including a one-time key in the the {\sc poll} also
ensures that the message is fresh (i.e., the key acts as a nonce).  


A communication neighbor $X \in \mathbb{N}_S$ that receives the {\sc poll}
stores its reception time $t_{SX}$, and extracts a random wait interval $T_X \in
[0,T_{max}]$ (Alg.~\ref{alg:exchange_neigh}, lines 2-5). After $T_X$ has
elapsed,  $X$ broadcasts a {\sc reply} message using a fresh MAC address, and
records the corresponding transmission time $t_X$
(Alg.~\ref{alg:exchange_neigh}, lines 6-10). The {\sc reply} contains encrypted
information for $S$, namely the signed neighbor identity, $Sig_X$, and the {\sc
poll} reception time: we refer to these data as $X$'s {\it commitment},
$\mathbb{c}_X$. The hash $h_{K'_{S}}$, derived from the verifier's public key,
$K'_{S}$, is also included to bind {\sc poll} and {\sc reply} belonging to the
same message exchange. 

Upon reception of a {\sc reply} message from a communication neighbor $Y$,
the verifier $S$ stores the reception time $t_{YS}$ and the
commitment $\mathbb{c}_Y$ (Alg.~\ref{alg:exchange_init}, lines 4-6).
A different communication neighbor of $S$, e.g., $X$, receives the {\sc reply}
message broadcast by $Y$,
if $Y$ is a communication  neighbor of both $S$ and $X$, i.e.,
$Y \in \mathbb{N}_S \cap \mathbb{N}_X$. In such case, $X$ too stores
the reception time $t_{YX}$ and the commitment $\mathbb{c}_Y$
(Alg.~\ref{alg:exchange_neigh}, lines 11-13).
Note that also {\sc reply} messages are anonymous, hence a node
records all  commitments it receives without knowing their origin.

After a time $T_{max}+\Delta+T_{jitter}$, $S$ broadcasts a {\sc reveal} message;
$\Delta$ accounts for the  propagation and contention lag of {\sc reply} messages
scheduled at time $T_{max}$, and  $T_{jitter}$ is a random time  added to thwart
jamming efforts on this message. Through the {\sc reveal}, the verifier $S$ (i) unveils its
identity  by including its signature and its public key to decrypt it, and (ii) proves
to be the author of the original {\sc poll}. The latter is achieved by
attaching  the encrypted hash $E_{k_{S}'}\{h_{K_{S}'}\}$
(Alg.~\ref{alg:exchange_init}, lines 7-9).

Once the identity of the verifier  is known, each neighbor $X$, which received
$S$'s original {\sc poll}, unicasts to $S$ an encrypted and signed {\sc report}
message containing its own position, the transmission time of its {\sc reply},
and the list of pairs of reception times and commitments  referring to the {\sc
reply} broadcasts it received (Alg.~\ref{alg:exchange_neigh}, lines 14-17).
Commitments are included `as they are', since only $S$ can decrypt  them and
match the identity of the nodes that created the commitments with  the reported
reception times.

\subsection{Position verification\label{subsec:verification}}

Once the message exchange is concluded, $S$ decrypts
the received data and acquires the position of all
neighbors that participated in the protocol, i.e.,
$\{p_X, \forall X \in \mathbb{N}_S\}$.
$S$ also knows the transmission time of its
{\sc poll} and learns the transmission time of all  subsequent {\sc reply}
messages, as well as the corresponding reception times recorded by
the recipients of such broadcasts.
Applying a ToF-based technique, $S$ can thus compute its
distance from each communication neighbor, as well as
the distances between
pairs of communication neighbors that happen to share a link.
In particular, denoting by $c$ the speed of
light, we define $d_{XY}=(t_{XY}-t_X) \cdot c$, i.e., the distance that $S$ computes
from the timing information it collected about the broadcast message sent by  $X$.
Similarly, we define  $d_{YX}=(t_{YX}-t_Y) \cdot c$, i.e., the distance that $S$ computes
using the information related to the broadcast by $Y$.
%
Exploiting its knowledge, the verifier  can run
verification tests to fill
the set $\mathbb{F}_S$ of faulty communication neighbors,
the set $\mathbb{V}_S$ of verified nodes, and the unverifiable set $\mathbb{U}_S$.

The first verification is carried through the {\bf Direct Symmetry (DS)} test,
detailed in Algorithm~\ref{alg:vt_ds}, where $|x|$ denotes the modulus of $x$ and
 $\left\|p_X-p_Y\right\|$ is the Euclidean distance between locations $p_X$ and $p_Y$.
For direct links between the verifier and each of its
communication neighbors, $S$ checks whether reciprocal ToF-derived distances
are consistent (i) with each other, (ii) with the position
advertised by the neighbor, and (iii) with a proximity range $R$.
The proximity range $R$ upper bounds the distance at which two nodes can
communicate, or, in other words, corresponds to the maximum nominal transmission range.

The first check is performed by comparing the distances $d_{SX}$ and $d_{XS}$
obtained from ranging, which shall not differ by more than twice the ranging
error  (Alg.~\ref{alg:vt_ds}, line 4). The second check verifies that the position advertised by the
neighbor is consistent with such distances,  within an error margin equal to
$2\epsilon_p+\epsilon_r$ (Alg.~\ref{alg:vt_ds}, line 5). This check is trivial but fundamental, since
it correlates positions to verified distances: without it, an attacker could fool
the verifier by simply advertising an arbitrary position along with correct
broadcast transmission and reception timings. Finally, $S$ verifies that
$d_{SX}$  is not larger than $R$ (Alg.~\ref{alg:vt_ds}, line 6), and declares a neighbor as faulty if a
mismatch surfaced in any of these checks\footnote{
The latter two checks are performed on both $d_{SX}$ and $d_{XS}$, however
in Algorithm~\ref{alg:vt_ds} they are done on $d_{SX}$ only, for clarity of presentation.}.

The {\bf DS} test implies  {\it direct} verifications
that compare trusted information collected by the verifier
against data advertised by each neighbor.
The content of the messages received by $S$, however, allows also
{\it cross}-verifications, i.e., checks on the information mutually
gathered by each pair of communicating neighbors. Such checks are done in
the {\bf Cross-Symmetry (CS)} test, in Algorithm~\ref{alg:vt_cs}.

The {\bf CS} test ignores nodes already declared as faulty by the {\bf DS} test
(Alg.~\ref{alg:vt_cs}, line 6) and only considers nodes that proved to be
communication neighbors between each other, i.e., for which ToF-derived mutual
distances are available (Alg.~\ref{alg:vt_cs}, line 7). Then, it verifies the
symmetry of such distances (Alg.~\ref{alg:vt_cs}, line 9), their consistency
with the positions declared  by the nodes (Alg.~\ref{alg:vt_cs}, line 10), and
their feasibility with respect to the proximity range  (Alg.~\ref{alg:vt_cs},
line 11). For each communication neighbor $X$, a link counter ${\sc l}_X$ and a
mismatch counter ${\sc m}_X$ are maintained. The former is incremented at every
new cross-verification on  $X$, and records the number of links between $X$ and
other communication neighbors of $S$ (Alg.~\ref{alg:vt_cs}, line 8). The latter
is incremented every time at least one of the cross-checks on distances and
positions fails (Alg.~\ref{alg:vt_cs}, line 12),  and identifies the potential
for  $X$ being faulty.

%

Once all neighbor pairs have been processed, a node $X$ is added
to the unverifiable set $\mathbb{U}_S$ if it shares less than two
neighbors with $S$
(Alg.~\ref{alg:vt_cs}, line 17). 
Indeed, in this case the information available on the
node is considered to be insufficient to tag the node as verified or
faulty (see Sec.~\ref{sec:analysis} for more details).
Otherwise, if $S$ and $X$ have two or more common neighbors, $X$ is declared as
faulty, unverifiable, or verified, depending on the percentage of mismatches
in the cross-checks it was involved
(Alg.~\ref{alg:vt_cs}, lines 18-22). More precisely, $X$ is added to $\mathbb{F}_S$,
$\mathbb{U}_S$ or $\mathbb{V}_S$, depending on whether
the ratio of the number of mismatches
to the number of checks is greater than, equal to, or less than
a threshold~$\delta$.

We point out that the lower the $\delta$, the
fewer the failed cross-checks needed to declare a node as faulty,
while the higher the $\delta$, the higher the probability of false negatives.
In the following, we set $\delta=0.5$ so that a majority rule is enforced:
the verifier makes a decision on the correctness of a node
by relying on the opinion of the majority of shared communication neighbors.
If not enough common neighbors
are available to build a reliable majority, the node is unverifiable.
As shown in the next section, this choice
makes our SNPD protocol robust to attacks in many different situations.

The third verification, the {\bf Multilateration (ML)} test, is
detailed in Algorithm~\ref{alg:vt_ml}.
The {\bf ML} test searches the verified set determined 
through the {\bf DS} and {\bf CS} algorithms
for suspicious situations, in which nodes in $\mathbb{V}_S$ declare a high
number of asymmetric links.
When a suspect node is found, the {\bf ML} test exploits as anchors
other nodes in $\mathbb{V}_S$, and multilaterates the actual position
of the node under verification.


The {\bf ML} test looks for each verified
neighbor $X$ of the initiator $S$ that did not notify
a link instead reported by another party $Y$ (Alg.~\ref{alg:vt_ml}, line 7).
When such a node is found, it is added to a {\it waiting set}
$\mathbb{W}_S$ (Alg.~\ref{alg:vt_ml}, line 8) and a curve $L_X(S,Y)$ is computed.
Such curve is the locus of points that can generate a transmission
whose Time Difference of Arrival (TDoA) at $S$ and $Y$ matches
that measured by the two nodes, i.e., $\left|t_{XS}-t_{XY}\right|$.
It is easy to verify that the curve is a hyperbola, which is
added to the set $\mathbb{L}_X$ (Alg.~\ref{alg:vt_ml}, line 9).

Once all couples of verified nodes have been checked, $\mathbb{W}_S$ is filled
with suspect neighbors. For each node $X$ in $\mathbb{W}_S$, $S$ exploits the
hyperbolae in $\mathbb{L}_X$ to multilaterate the position of $X$, referred to
as $p_X^{ML}$, similarly to what is done in~\cite{capkun:secpos} 
(Alg.~\ref{alg:vt_ml}, line 14). Note that  $\mathbb{L}_X$ must include at least
two hyperbolae for $S$ to be able to compute the position $X$ through 
multilateration, and this implies the presence of at least two shared neighbors
between $S$ and $X$ (Alg.~\ref{alg:vt_ml}, line 13). The resulting position
$p_X^{ML}$ is then compared against that advertised by $X$, $p_X$. If the
difference exceeds a given error margin, neighbor $X$ is moved from the verified
set to the faulty one (Alg.~\ref{alg:vt_ml}, lines 15-17).

\section{Security analysis}
\label{sec:analysis}

We analyze the security properties of the proposed
scheme in presence of adversarial nodes, whose objective is to make
the verifier believe that the fake positions they advertise are correct.
We consider scenarios of increasing complexity: we start by discussing
the basic workings of the SNPD protocol
in presence of a single adversary and different shared neighborhoods;
we then move to the case of multiple adversaries, at first assuming they act
independently and, then, that they cooperate to perform the attack;
finally, we examine the resilience of the scheme to a number of well-known attacks.


\subsection{Single adversary, no common neighbors}
\label{subsec:1a0n}

Consider a verifier $S$ that starts the SNPD protocol in presence
of an adversary $M$, with which it shares no common neighbor.
In order to bring a successful attack, $M$ must
tamper with the data $S$ uses for ranging,
so that the resulting distance confirms its fake advertised position.
To this end, $M$ can forge at its convenience the time information
in the messages it generates. In particular, let  $p'_M$ be
the fake position that $M$ wants to advertise; we denote
by $t'_{SM}$ the fake timing that $M$ introduces in its {\sc reply},
and by $t'_M$ the fake timing inserted in its {\sc report} (in addition
to $p'_M$).

The {\bf DS} test (Alg.~\ref{alg:vt_ds}) run
by $S$ on $M$ checks the consistency between distances, by verifying that
$\left|d_{SM}-d_{MS}\right|\leq 2\epsilon_r$, or:
\begin{equation}
\left| (t'_{SM}-t_S)\cdot c - (t_{MS}-t'_M)\cdot c \right| \leq 2\epsilon_r
\label{eq:vt1a}
\end{equation}
and that positions are also coherent with
the distances, i.e.,
$\left| \left\|p_S-p'_M\right\| - d_{SM}\right| \leq 2\epsilon_p+\epsilon_r$, or,
equivalently:
\begin{equation}
\left| \left\|p_S-p'_M\right\| - (t'_{SM}-t_S)\cdot c \right| \leq 2\epsilon_p+\epsilon_r
\label{eq:vt2}
\end{equation}
Thus, the adversary must forge $t'_M$ and $t'_{SM}$, so that
(\ref{eq:vt1a})--(\ref{eq:vt2}) still hold
after its real position $p_M$ is replaced with $p'_M$.
Solving the equation system obtained by setting
the error margin to zero
in (\ref{eq:vt1a})--(\ref{eq:vt2}),
we obtain:
\begin{equation}
\label{eq:fake_tM}
t'_M =  t_{MS} - \frac{\left\|p_S-p'_M\right\|}{c}
     =  t_M + \frac{\left\|p_S-p_M\right\|}{c} - \frac{\left\|p_S-p'_M\right\|}{c}
\end{equation}
\begin{equation}
\label{eq:fake_tSM}
t'_{SM}  =  t_S + \frac{\left\|p_S-p'_M\right\|}{c}
        =  t_{SM} - \frac{\left\|p_S-p_M\right\|}{c} + \frac{\left\|p_S-p'_M\right\|}{c}
\end{equation}
Note that $p'_M$ is chosen by $M$, and that $M$ knows
$t_{M}$ in (\ref{eq:fake_tM}) (since this is the actual transmission
time of its own {\sc reply}) and
$t_{SM}$ in (\ref{eq:fake_tSM}) (since this is
the time at which it actually received the {\sc poll} from $S$).
We therefore have a system of two equations that $M$ can solve, in the
two unknowns $t'_M$ and $t'_{SM}$, only if it is aware of $p_S$,
i.e., it is a knowledgeable adversary.
We stress that, for $M$ to be knowledgeable, two conditions must hold:
first, $M$ must have previously run the SNPD protocol to discover
the identity and position of its neighbors; second,
the verifier's position must have not changed since such discovery procedure.
Clearly, as $M$ cannot foresee when $S$ starts the SNPD protocol, such conditions are extremely
hard to fulfill, especially in a highly dynamic environment such as the vehicular one.


Nevertheless, if $M$ is aware of $S$'s location, the advertised position
$p'_M$ will pass the {\bf DS} test provided that it is
within the proximity range $R$, as shown in Fig.~\ref{fig:security_1a0n}.
Given such potential weakness,
the SNPD protocol marks isolated neighbors as unverifiable in the {\bf CS} test,
even if they pass the {\bf DS} test.

\subsection{Single adversary, one common neighbor}
\label{subsec:1a1n}

We now add to the previous scenario a node $X$, which is a correct
neighbor, common to $S$ and $M$.
Recall that, in bringing
its attack, $M$ can forge messages with altered information,
but it cannot modify the content of messages sent by other nodes,
since they are all encrypted and signed.

The discussion in Sec.~\ref{subsec:1a0n}
applies again, since the fake position advertised by $M$
needs to pass the {\bf DS} test: $M$ must be aware of  $S$'s current position and
must forge $t'_{M}$ and $t'_{SM}$ according to $p_S$ and $p'_{M}$.
However, the presence of the common neighbor introduces two additional
levels of security.

First, the {\sc poll} and {\sc reply} messages
are anonymous, hence $M$ does not know if the verifier is $S$ or $X$
upon reception of such messages.
However, if it wants to take part in the protocol, $M$ is forced
to advertise the fake {\sc poll} reception time $t'_{SM}$ in its
{\sc reply} message, before receiving the {\sc reveal} and discovering
the verifier's identity.
The only option for $M$ is then to randomly guess who the
verifier is, and properly change $t_{SM}$ into $t'_{SM}$, as
in (\ref{eq:fake_tSM}), and this implies a 0.5 probability of
failure in the attack.

Second, the {\bf CS} test on the pair $(M,X)$ requires that
$\left|d_{XM}-d_{MX}\right|\leq 2\epsilon_r$ and
$\left| \left\|p_X-p_M\right\| - d_{XM}\right| \leq 2\epsilon_p+\epsilon_r$.
Exactly as before, to pass these checks, $M$ is forced to advertise the fake timings:
\begin{eqnarray}
\label{eq:fake_tM2}
t'_M    & = & t_M + \frac{\left\|p_X-p_M\right\|}{c} - \frac{\left\|p_X-p'_M\right\|}{c}\\
\label{eq:fake_tXM}
t'_{XM} & = & t_{XM} - \frac{\left\|p_X-p_M\right\|}{c} + \frac{\left\|p_X-p'_M\right\|}{c}
\end{eqnarray}
If $M$ knows $X$'s current position $p_X$, it can solve (\ref{eq:fake_tXM})
and announce the forged $t'_{XM}$ in its {\sc report} to $S$.
However, (\ref{eq:fake_tM2}) introduces a second expression for $t'_M$,
whereas $M$ can only advertise one single $t'_M$. In order to pass
both {\bf DS} and {\bf CS} tests, $M$ needs to announce a $t'_M$ that satisfies (\ref{eq:fake_tM})
and (\ref{eq:fake_tM2}), which implies:
\begin{equation}
\label{eq:conditionA_1a1n}
\left\|p_S-p_M\right\| - \left\|p_S-p'_M\right\| =
\left\|p_X-p_M\right\| - \left\|p_X-p'_M\right\|
\end{equation}

In other words, $M$ is constrained to choose locations with the
same distance increment (or decrement) from
$S$ and $X$. In~(\ref{eq:conditionA_1a1n}),
$p_S$, $p_X$, and $p_M$ are fixed and known,
hence distances between $p_S$ and $p_M$, and between $p_X$ and $p_M$
can be considered as constant. Since $p'_M$ is variable
over the plane, we rewrite (\ref{eq:conditionA_1a1n}) as
$\left\|p_X-p'_M\right\| - \left\|p_S-p'_M\right\| = k$,
which is the equation describing a hyperbola with foci in $p_S$ and $p_X$,
and passing through $p_M$.
It follows that only positions on such hyperbola satisfy the
four constraints in  (\ref{eq:fake_tM}), (\ref{eq:fake_tSM}),
(\ref{eq:fake_tM2}), and (\ref{eq:fake_tXM}), and
$p'_M$
must lie on that curve in order to pass all tests.
Examples of this condition are shown in Fig.~\ref{fig:security_1a1n}.

Summarizing, the presence of a common neighbor $X$ drastically reduces
the vulnerability of the verifier to attacks, since $M$ is
now required (i) to be knowledgeable,
(ii) to correctly guess the verifier's identity,
and (iii) to advertise a fake position only along a specific curve.
However, since some space for successful attacks remains,
the {\bf CS} test marks as unverifiable nodes that passed the {\bf DS} test
but share only one neighbor with the verifier.
We also stress that, if $M$ tweaks the timings so as to pass
the {\bf DS} test and does not care about the matching with $X$, it will still be
tagged as unverifiable.

\subsection{Single adversary, two or more common neighbors}
\label{subsec:1a2+n}

In the case of two or more common neighbors, we split the discussion
into the two following cases: (i) a generic network topology and (ii)
collinear nodes. 

{\em (i) Generic network topology}.
When a second correct neighbor $Y$ is shared
between $S$ and $M$~\footnote{Note that we do not make any assumption
on the connectivity between $X$ and $Y$.}, the discussion in Sec.~\ref{subsec:1a1n}
can be extended as follows. We noting that, as before, 
the adversary $M$ has to be knowledgeable,
but a
second common neighbor reduces to 0.33 the probability that $M$ correctly
guesses the verifier's identity. More importantly, by applying the same
reasoning as in Sec.~\ref{subsec:1a1n}, $M$ has now to forge four time values,
i.e., $t'_M$, $t'_{SM}$, $t'_{XM}$, and $t'_{YM}$, so that six equations are
satisfied, i.e., (\ref{eq:fake_tM}), (\ref{eq:fake_tSM}), (\ref{eq:fake_tM2}),
(\ref{eq:fake_tXM}), and the two equations corresponding to the cross-check
with the second common  neighbor $Y$~\footnote{The latter  two equations can be
obtained from (\ref{eq:fake_tM2})--(\ref{eq:fake_tXM})  by replacing  $p_X$,
$t_{XM}$ and $t'_{XM}$, respectively,  with $p_Y$, $t_{YM}$ and  $t'_{YM}$.}.

To fulfill the constraints on $t'_M$, now $M$  has
to announce a position $p'_M$ that is equally farther from
(or closer to) $S$, $X$ and $Y$ with respect to its actual
location $p_M$.
The point satisfying such condition lies at the
intersection of three hyperbolae with foci in $p_S$ and $p_X$,
$p_S$ and $p_Y$, $p_X$ and $p_Y$, respectively,
and such single point actually corresponds to the real position of the adversary, $p_M$.

Accordingly, in presence of two common neighbors,
the {\bf CS} test marks a node with no mismatches as verified.
The majority rule (i.e., $\delta=0.5$) results
instead in the adversary being tagged as faulty
when mismatches are recorded with both common neighbors.
Finally, the adversary is added to the unverifiable set if
it is capable of fooling $S$ and either $X$ or $Y$, since that leads
to one mismatch over two links checked.

We stress that deceiving $S$ and one of the common neighbors
requires, beside the knowledge of their current positions and a correct guess
on the verifier's identity, also the pinning of which {\sc reply}
comes from which neighbor
(i.e., $M$ must randomly map $t_{XM}$ onto $p_X$ and
$t_{YM}$ onto $p_Y$ for the computations on the hyperbolae to work).
Thus, the guess taken by $M$ in the hope of being
marked as unverifiable has a success probability of 0.165, jointly given by
the probability of guessing the right
verifier (0.33) and the probability of guessing the right
mapping (0.5) of {\sc reply} reception times onto neighbor positions.

When three or more common neighbors are present between $S$ and $M$,
the chances of a successful attack drop to zero.
Indeed, not only the probability of guessing
the right originators of the different messages shrinks as
the size of the common neighborhood grows, but the majority rule
dooms the adversary to insertion in the faulty set,
even when all random guesses are exact.
By extending the above analysis on the hyperbolae, we observe that,
with a threshold $\delta=0.5$, when $S$ and $M$ share $n \geq 3$
communication neighbors, the mismatch-to-links ratio is $\frac{n-1}{n} > \delta$.

A summary of the security of the SNPD protocol, in presence of a single
adversary and in a generic network topology, is presented in
Tab.~\ref{tab:security_1a}, where different rows identify different behaviors of
the neighbor $X$ under verification by  $S$. The columns represent the number of
correct neighbors shared by $S$ and $X$. For each combination, we report the set
to which $X$ is assigned by $S$, possibly with a probability value due to the
adversary's random guessing on the roles of neighbors.

{\em (ii) Collinear nodes}.
When the majority of common neighbors
is collinear to $S$ and an adversary $M$, and lies on the same side as $S$ with
respect to $p_M$, a degree of freedom exists for the attacker.
Indeed, $M$ is verified if it announces a fake position that is
collinear with $p_M$ and $p_S$, within a distance $R$ from
$S$, and such that the majority of the common neighbors still
lies on the same side as $S$ with respect to $p'_M$.
This case, however, hardly leads to an advantage for the adversary, since $p'_M$
must remain aligned with the positions of the other nodes, must respect the ordering
with the majority of them, and cannot exceed $S$'s proximity range.

\subsection{Multiple independent adversaries}
\label{subsec:ma}

We now consider the presence of multiple uncoordinated adversaries.
It is easy to see that independent attackers
damage each other, by announcing false positions that
reciprocally spoil the time computations discussed in the
previous sections.
Cross checks on couples
of non-colluding adversaries will always result in mismatches
in the {{\bf CS} test, increasing the chances that such nodes are tagged as
faulty by the initiator.

Where multiple independent attackers can harm the system is in
the verification of correct neighbors. As a matter of fact, a
node is ruled verified if it passes the strict majority
of cross controls it undergoes. A correct node surrounded by
several adversarial neighbors could thus be marked as faulty (unverifiable),
if it shares  with the initiator a number of adversarial nodes greater than (equal to)
the number of correct nodes.
An example is provided in Fig.~\ref{fig:security_ma_independent}.
%
However, it is to be said that, under the assumption that the percentage of attackers
among all nodes in the network is small, situations where a correct
node shares mostly uncoordinated adversarial neighbors with the initiator
are very unlikely to occur.

\subsection{Multiple colluding adversaries, basic attack}
\label{subsec:ma_colluding_basic}

Coordinated attacks carried out by colluding adversaries
are obviously harder to counter than those independently
led by individual adversarial nodes. The SNPD protocol is
resistant to coordinated attacks, unless the presence of
colluding adversaries in the neighborhood of the initiator
node is overwhelming.

The goal of adversarial nodes remains that of inducing the initiator $S$
into trusting the fake positions they announce. The basic way they can
cooperate to that end is by mutually validating the false information they
generate. Indeed, colluding adversaries can advertise to $S$
reception times (of reciprocal {\sc reply} messages) forged so that
the values derived through ToF-based ranging confirm
the positions they made up in the {\bf CS} test.
In other words, a perfect cooperation results in the colluding
adversaries' capability of ``moving'' all links among them without
being noticed by the initiator.
Our SNPD protocol can counter the basic attack from colluders, as long
as 50\% {\em plus one} of the neighbors in common to the verifier and an
adversary are correct. Indeed, a strict majority of correct shared
neighbors allows the identification of attackers through the {\bf CS} test.
An example with three colluding attackers
is provided in Fig.~\ref{fig:security_ma_colluding}.

\subsection{Multiple colluding adversaries, hyperbolae-based attack}
\label{sec:ma_colluding_hyperbola}

A more sophisticated version of the basic coordinated attack can be
organized by colluding adversaries as follows.
Having received the {\sc poll} message, the
attackers not only agree on the identity of the initiator $S$, but also
pick a common neighbor $X$ that they share with $S$: each colluder
determines the hyperbola with foci $S$, $X$, and passing through its
own actual position, and announces a fake position on such curve.
This allows the adversaries to announce correct links (i) with the
initiator $S$, (ii) with the selected neighbor $X$, and (iii) among
themselves. Node $X$ becomes an involuntary allied in the attack:
in order to work properly, the {\bf CS} test, based on the majority rule,
needs that more than
50\% {\it plus three} of the common neighbors between the initiator
and communicating node are correct. The two additional correct
neighbors are required to counter the effect of $X$ becoming an
unintentional colluder during the cross verification.

\subsection{Multiple colluding adversaries, {\sc reply}-disregard attack}
\label{subsec:ma_colluding_disregard}

A second variation to the attack presented in
Sec.~\ref{subsec:ma_colluding_basic} relies on a coordinated action
against {\sc reply} messages received from correct nodes.
As a matter of fact, the {\bf CS} test can control the symmetry
of links between couples of neighbors only if ToF-based ranging
is performed in both directions. Thus, by intentionally excluding
from their {\sc report}  the commitments received from
correct nodes while including all those received by colluding nodes,
adversaries can selectively avoid cross symmetry tests with correct
nodes, so that no mismatches are found. We refer to this as a
{\sc reply}-disregard attack and stress that it requires at least
three colluding nodes forming a clique,
or the adversaries would result unverifiable to the initiator, since
they would share less than two (bidirectional) neighbors with it.

The SNPD protocol is robust to {\sc reply}-disregard attacks, thanks to the
controls run in the {\bf ML} test. More precisely, an adversary carrying out a
disregard attack together with $N$ colluders can safely advertise up to $N-1$
wrong reception times from correct nodes, being still tagged as verified by the
majority rule. This means that there must be at least $N+1$ correct neighbors,
shared by an adversary and the initiator, for the adversary to be forced to
disregard one or more {\sc reply}, and for two correct shared neighbors to be in
the condition of participating in the {\bf ML} test and identify the colluder.
This means that 50\% {\em plus two} of the shared neighbors must be correct
for our SNPD protocol to work properly.

As a final remark on coordinated attacks, we comment on the significant
resources and a strong effort they require from the colluding adversaries.
Colluders have to share out-of-band links through which
they can exchange information to coordinate the attack, upon
reception of the {\sc poll} message. Exploiting such links,
they first have to agree on the initiator's identity, either
by a shared random guess or by employing a multilateration
technique to disclose it.
Then, colluders have to inform each other about the fake positions they
will announce, and about the estimated transmission time of their
{\sc reply} messages: this way, each cooperating adversary is able
to recognize the anonymous {\sc reply} of a colluder node and to compute a
reception time that is consistent with the fake position advertised
by such colluder. Finally, this exchange of information must occur in a
very limited time interval after the {\sc poll} message has been
broadcast, so that colluders can transmit their {\sc reply} messages
well before the $T_{max}$ deadline.

\subsection{Denial of Service (DoS) attacks}\label{subsec:dos}

{\bf Jamming.} An adversary $M$ may jam the channel and erase {\sc reply} or
{\sc report} messages. To successfully perform such an attack, $M$ should jam
the medium continuously for a long time, since it cannot know when exactly each
of the nodes will transmit its {\sc reply} or {\sc report} message. Or, $M$
could erase the {\sc reveal} message, but, again, jamming should cover the
entire $T_{jitter}$ time; jamming a specific  {\sc reply} transmission is not
straightforward either as the {\sc reply} transmission time is randomly chosen
by each node. Overall, there is no easy point to target; a jammer has to
basically jam throughout the SNPD execution, an action that is possible for any
wireless protocol and orthogonal to our problem.

{\bf Clogging.} An adversary could induce SNPD traffic in an attempt to
congest the wireless channel, e.g., by initiating the protocol multiple times in
a short period and getting repeated {\sc reply} and {\sc report} messages from
other nodes. 
{\sc report} messages are  large and unicast, and generated in a
short period after the reception of the {\sc reveal} message. 
They are thus likely to cause the most damage. 
However, SNPD has a way of preventing that: 
the initiator must unveil its identity before such messages are
transmitted by neighbors. An exceedingly frequent initiator can be
identified and rate-limited, its excessive {\sc reveal} messages ignored.
Conversely, {\sc reply} messages are small in size, they are broadcast
(and thus require no ACK) and they are
spread over the time interval $T_{max}$. Their damage is somewhat limited,
but their unnecessary transmission is much harder to thwart.
Indeed, {\sc reply} messages should be sent following an anonymous
{\sc poll} message; such anonymity is a requirement that is hard to dismiss,
since it is instrumental to keeping adversaries unknowledgeable.  
As a general rule, correct nodes can reasonably
self-limit their responses if {\sc poll}s arrive at excessive rates. Overall,
clogging DoS have only local effect, within the neighborhood of the adversary,
which could anyway resort to jamming and obtain the same effect.



\subsection{Adversarial use of directional antennas} 

Assume that adversarial nodes are equipped with directional antennas and
multiple radio interfaces. Then, as a correct node $S$ starts the SNPD protocol,
a knowledgeable adversary $M$ can send {\sc reply} messages through the
different interfaces at different time instants, so as to fool the communication
neighbors shared by $M$ and $S$: a correct neighbor $X$ would record a time
$t'_{MX}$, which is compliant with the fake position, $p'_M$, announced by $M$
and, thus, can pass the corresponding cross check in the {\bf CS} test. If the
adversary is able to fool a sufficient number of neighbors, it succeeds and is
tagged as verified; however, we stress that the adversary needs as many
directional antennas and radio interfaces as the number of neighbors it wants to
fool. Moreover, it must hope that no two such neighbors are within the beam of
the same antenna. The complexity, cost, and chances of failure make this attack
hardly viable.

\section{Performance evaluation}
\label{sec:performance}

To test our SNPD protocol, we selected a real-world road topology
that consists of a 5$\times$5 km$^2$ portion of the urban
area of the city of Zurich~\cite{zurichtraces}. These traces describe
the individual movement of cars through a queue-based model
calibrated on real data:
they thus provide a realistic representation of vehicular mobility at
both microscopic and macroscopic levels.
We extracted 3 hours of vehicular
mobility, in presence of mild to heavy traffic density conditions;
the average number
of cars in the area at a given time is 1200.

Traces have a time discretization of 1 s. Thus, given a trace, every second
we randomly  select 1\% of the nodes
as  verifiers. For each node, we consider that all devices  within the proximity
range $R$ are communication neighbors of the node.
Clearly, the larger the $R$, the higher the number of neighbors taking part
in the same instance of the SNPD protocol: for example
for $R$ equal to 50~m and 500~m,
the average node degree is 8 and 104.8 and the variance is 5.9 and 71.8,
respectively.
Also, we set $\epsilon_r$ to 6.8~m and
$\epsilon_p$  to 5~m~\cite{techrep}.

Since {\em unknowledgeable} adversaries are always tagged as faulty
in the {\bf DS} test, in the following we present results considering
that all adversaries are always {\em knowledgeable}. We stress that this is
a very hard condition to meet in dynamic networks, hence all results
are to be considered as an upper bound to the success probability of
an attack.

When independent adversaries are
considered, we randomly select a ratio (a varying parameter in our
analysis) of the  nodes as attackers. In case of colluders, instead, we
randomly select some nodes as adversaries, and for each we further randomly
identify neighbors who will collude with it so as to form an attackers group of
size $\sigma$ (or up to the number of neighbors available).
We assume that colluding adversaries perform hyperbolae-based
attacks, which, as previously discussed, are the hardest to contrast.
For every scenario under
study,  we statistically quantify the outcome of the verification test and
compare it to the actual behavioral model of the nodes (namely, correct or
adversary).

We first report results in terms of probabilities
that the tests return false positives and false negatives
(Figs.~\ref{fig:zurich_malw} and~\ref{fig:zurich_Rw}) as well as of probability
that a (correct or adversary) node is tagged as unverifiable
(Figs.~\ref{fig:zurich_malu} and~\ref{fig:zurich_Ru}). The former gauge the
reliability of our scheme, while the latter is a mark of the protocol accuracy.
The plots showing the false positives and false negatives, when  the ratio of
adversaries varies and $R$=250~m, confirm that our scheme errs on the side of
caution: indeed, as the number of adversaries increases, it is more likely for a
correct node to be mislabeled than for an adversary  to be verified (the latter
probability amounting to less than 0.02). Instead, widening the proximity range
with a fixed adversary ratio, namely 0.05,
only plays into the verifier's hands, thanks to the
greater number of nodes (the majority of which are correct) that can be tested.
As for the probability that a node is unverifiable, while little sensitivity to
the ratio of adversaries is observed, a small $R$ (hence fewer neighbors) affects
the protocol capability to reach a conclusive verdict on either correct or
adversary nodes. We also estimated that  the degree of freedom that a successful
adversary has in setting its fake position, for $R$=250 m and a ratio of 0.05
attackers, is such that, on average, the fake and actual positions of a verified
adversary are collinear and differ by 40~m.

We then fix the adversaries ratio to 0.05 and $R$ to 250~m
and we consider the presence of colluders.
Figs.~\ref{fig:coll_zurich_w}~and~\ref{fig:coll_zurich_u} show the excellent
performance of our scheme as the colluder group size $\sigma$ varies.
The impact of colluders on the results appears to be negligible,
mainly thanks to the large number of neighbors
defeating even big groups of colluders.

Finally, we comment on the overhead introduced by SNPD, in terms of number and
size of messages. SNPD generates at most $2N+2$ messages for one execution
initiated by a verifier with $N$ communication neighbors.  This is twice the
cost of an unsecured NPD protocol that would consist of one poll and $N$
position replies from neighbors. Moreover, SNPD messages are relatively small in
size: with SHA-1 hashing and ECDSA-160 encryption~\cite{ieee1363}, 
the length of
signatures is 21 bytes (with coordinates compression). Assuming that messages
include headers with 4-byte source and destination identifiers and 1-byte
message type field, {\sc POLL}, {\sc REPLY}, and {\sc REVEAL} are all less than
100 bytes in size (to be precise, 26, 71, and 67 bytes, respectively). The {\sc
REPORT} length is variable, depending on the number of commitments it carries:
e.g., for 5 commitments, its size is only 295 bytes, and up to 28 commitments
can fit in a single 1500-byte IP packet.  
Obviously, the on-demand nature of
the protocol makes it best suited to event-triggered
applications, such as safety and tolling ones. In these scenarios, 
SNPD induces very low
overhead in the network. The limited number and the small size of messages 
make
the proactive use of the protocol feasible, for relatively low rate execution,
e.g., once in a few tens of seconds.

\section{Conclusion}

We proposed a lightweight, distributed scheme for securely discovering the
position of communication neighbors in vehicular ad hoc networks. Our solution does
not require the use of a-priori trustworthy nodes, but it leverages the
information exchange between neighbors. Our
analysis showed the scheme to be very effective in identifying independent
as well as colluding adversaries. Results derived using realistic vehicular
traces confirmed such ability and highlighted the good performance of
our solution in terms of both false negatives/positives
and uncertain neighbor classifications.

Future work will aim at assessing the performance of the proposed secure
neighbor position discovery protocol when adversaries  have partial or
out-of-date knowledge on the other nodes' positions, and at adapting our scheme
to a high-frequency proactive utilization.

\newpage

\begin{table*}[htbp]
 \centering
\caption{\label{tab:notation} Summary of notations}
  \begin{tabular}{ll}
    \hline 
    {\textit{Notation}} & {\textit{Description}} \\
    \hline 
    {$k_X$ (resp. $K_X$)} & {private (resp. public) key of node $X$} \\
    {$k'_X$ (resp. $K'_X$)} & {private (resp. public) one-time key of node $X$} \\
    {$t_X$ (resp. $t'_X$)} & {actual (resp. fake) transmission time of a message by node $X$} \\
    {$t_{XY}$  (resp. $t'_{XY}$)} & {actual (resp. fake) reception time at node $Y$ of a message sent by node $X$} \\
    {$p_X$ (resp. $p'_X$)} & {actual (resp. fake) position of node $X$} \\
    {$d_{XY}$} & {distance between nodes $X$ and $Y$} \\
    {$\epsilon_p$ (resp. $\epsilon_r$)} & {position (resp. ranging) error} \\
    {$R$} & {node proximity range} \\
    
    {$\mathbb{N}_X$} & {current set of communication neighbors of node $X$} \\
    {$T_X$} & {random wait interval after reception of {\sc poll} at node $X$} \\    
    {$Sig_X$} & {signed identity of node $X$} \\
    {$\mathbb{c}_X$} & {{\em commitment} of node $X$} \\
    {$\mathbb{V}_X$} & {set of {\em verified} communication neighbors of node $X$} \\
    {$\mathbb{U}_X$} & {set of {\em unverifiable} communication neighbors of node $X$} \\
    {$\mathbb{F}_X$} & {set of {\em faulty} communication neighbors of node $X$} \\   
    \hline
  \end{tabular}
\end{table*}

\restylealgo{algoruled}
\SetAlgoSkip{medskip}
\incmargin{2pt}
\dontprintsemicolon
\linesnumbered
\begin{algorithm}[h]
\SetKwFor{Node}{node}{do}{end}
\SetKwFor{After}{after}{do}{end}
\SetKwFor{When}{when}{do}{end}
\SetLine
\Node{$S$}{
	$S \rightarrow *$ : $\langle \text{\sc poll},K_{S}' \rangle$\;
	$S$ : store $t_S$\;
	\When{receive {\sc reply} from $Y \in \mathbb{N}_S$}{
		$S$ : store $t_{YS},\mathbb{c}_Y$\;
	}
	\After{$T_{max}+\Delta+T_{jitter}$}{
		$S \rightarrow *$ : $\langle \text{\sc reveal},E_{k_{S}'}\{h_{K_{S}'}\},K_S,Sig_S \rangle$\;
	}
}
\caption{Message exchange protocol: verifier node}
\label{alg:exchange_init}
\end{algorithm}
\decmargin{2pt}
\restylealgo{algoruled}
\SetAlgoSkip{medskip}
\incmargin{2pt}
\dontprintsemicolon
\linesnumbered
\begin{algorithm}[h]
\SetKwFor{Node}{node}{do}{end}
\SetKwFor{After}{after}{do}{end}
\SetKwFor{When}{when}{do}{end}
\SetLine
\ForAll{$X \in \mathbb{N}_S$}{
	\When{receive {\sc poll} by $S$} {
		$X$ : store $t_{SX}$\;
		$X$ : extract $T_X$ uniform r.v. $\in [0, T_{max}]$
	}
	\After{$T_X$}{
		$X$ : $\mathbb{c}_X=E_{K_{S}'}\{t_{SX},K_X,Sig_X\}$\;
		$X \rightarrow *$ : $\langle \text{\sc reply},\mathbb{c}_X, h_{K'_S} \rangle$\;
		$X$ : store $t_X$\;
	}
	\When{receive {\sc reply} from $Y \in \mathbb{N}_S \cap \mathbb{N}_X$}{
		$X$ : store $t_{YX},\mathbb{c}_Y$\;
	}
	\When{receive {\sc reveal} from $S$}{
		$X$ : $\mathbb{t}_X=\{(t_{YX},\mathbb{c}_Y)\:\, \forall\, Y\, \in \mathbb{N}_S \cap \mathbb{N}_X\}$\;
		$X \rightarrow S$ : $\langle \text{\sc report},E_{K_{S}}\{p_X,t_X,\mathbb{t}_X,Sig_X\} \rangle$\;
	}
}
\caption{Message exchange protocol: neighbor node}
\label{alg:exchange_neigh}
\end{algorithm}
\decmargin{2pt}

\restylealgo{algoruled}
\SetAlgoSkip{medskip}
\incmargin{2pt}
\dontprintsemicolon
\linesnumbered
\begin{algorithm}[h]
\SetKwFor{Node}{node}{do}{end}
\SetKwIF{orIfa}{orElseIfa}{orElsea}{if}{or}{else if}{else}{endif}
\SetKwIF{orIfb}{orElseIfb}{orElseb}{\hspace*{6pt}}{or}{else if}{else}{endif}
\SetKwIF{orIfc}{orElseIfc}{orElsec}{\hspace*{6pt}}{then}{else if}{else}{endif}
\SetLine
\Node{$S$}{
	$S$ : $\mathbb{F}_S \leftarrow \emptyset$\;
	\ForAll{$X \in \mathbb{N}_S$}{
		\lorIfa{$\left|d_{SX} - d_{XS}\right| > 2 \epsilon_r$}\;
		\lorIfb{$\left|\left\|p_S-p_X\right\| - d_{SX}\right| > 2\epsilon_p + \epsilon_r$}\;
		\orIfc{$d_{SX} > R$}{
			$S$ : $\mathbb{F}_S \leftarrow X$\;
		}
	}
}
\caption{Direct Symmetry (DS) test}
\label{alg:vt_ds}
\end{algorithm}
\decmargin{2pt}
\vskip -5mm

\restylealgo{algoruled}
\SetAlgoSkip{medskip}
\incmargin{2pt}
\dontprintsemicolon
\linesnumbered
\begin{algorithm}[h]
\SetKwFor{Node}{node}{do}{end}
\SetKwIF{orIfa}{orElseIfa}{orElsea}{if}{or}{else if}{else}{endif}
\SetKwIF{orIfb}{orElseIfb}{orElseb}{\hspace*{6pt}}{or}{else if}{else}{endif}
\SetKwIF{orIfc}{orElseIfc}{orElsec}{\hspace*{6pt}}{then}{else if}{else}{endif}
\SetLine
\Node{$S$}{
	$S$ : $\mathbb{U}_S \leftarrow \emptyset$, $\mathbb{V}_S \leftarrow \emptyset$\;
	\ForAll{$X \in \mathbb{N}_S$, $X \notin \mathbb{F}_S$}{
		$S$ : $l_X=0$, $m_X=0$\;
	}
	\ForAll{$\left(X,Y\right) \:|\: X,Y \in \mathbb{N}_S$, $X,Y \notin \mathbb{F}_S$, $X \neq Y$}{
		\If{$\exists\; d_{XY},d_{YX}$}{
			$S$ : $l_X=l_X+1$, $l_Y=l_Y+1$\;
			\lorIfa{$\left|d_{XY} - d_{YX}\right| > 2\epsilon_r$}\;
			\lorIfb{$\left|\left\|p_X-p_Y\right\| - d_{XY}\right| > 2\epsilon_p+\epsilon_r$}\;
			\orIfc{$d_{XY} > R$}{
				$S$ : $m_X=m_X+1$, $m_Y=m_Y+1$\;
			}
		}
	}
	\ForAll{$X \in \mathbb{N}_S$, $X \notin \mathbb{F}_S$}{
		\lIf{$l_X < 2$}{
			$S$ : $\mathbb{U}_S \leftarrow X$\;
		}
		\lElse{\Switch{$\frac{m_X}{l_X}$}{
			\lCase{$\frac{m_X}{l_X} > \delta$}{
				$S$ : $\mathbb{F}_S \leftarrow X$\;
			}
			\lCase{$\frac{m_X}{l_X} = \delta$}{
				$S$ : $\mathbb{U}_S \leftarrow X$\;
			}
			\lCase{$\frac{m_X}{l_X} < \delta$}{
				$S$ : $\mathbb{V}_S \leftarrow X$\;
			}
		}}
  }
}
\caption{Cross-Symmetry (CS) test}
\label{alg:vt_cs}
\end{algorithm}
\decmargin{2pt}

\restylealgo{algoruled}
\SetAlgoSkip{medskip}
\incmargin{2pt}
\dontprintsemicolon
\linesnumbered
\begin{algorithm}[h]
\SetKwFor{Node}{node}{do}{end}
\SetKwIF{orIfa}{orElseIfa}{orElsea}{if}{or}{else if}{else}{endif}
\SetKwIF{orIfb}{orElseIfb}{orElseb}{\hspace*{6pt}}{or}{else if}{else}{endif}
\SetKwIF{orIfc}{orElseIfc}{orElsec}{\hspace*{6pt}}{then}{else if}{else}{endif}
\SetLine
\Node{$S$}{
	$S$ : $\mathbb{W}_S \leftarrow \emptyset$\;
	\ForAll{$X \in \mathbb{V}_S$}{
		$S$ : $\mathbbm{L}_X \leftarrow \emptyset$\;
	}
	\ForAll{$\left(X,Y\right) \:|\: X,Y \in \mathbb{V}_S$, $X \neq Y$}{
		\If{$\exists\; t_{XY}$  \mbox{\rm and} $\nexists\; t_{YX}$}{
			\lIf{$X \notin \mathbb{W}_S$}{
				$S$ : $\mathbb{W}_S \leftarrow X$\;
			}
			$S$ : $\mathbb{L}_X \leftarrow L_X(S,Y)$\;
		}
	}
	\ForAll{$X \in \mathbb{W}_S$}{
		\If{$\left|\mathbb{L}_X\right| \geq 2$}{
			$S$ : $p^{ML}_X = \arg\min_p\sum_{L_i,L_j\in\mathbb{L}_X } \left\|p- L_i\cap L_j\right\|^2$\;
			\If{$\left\|p_X - p^{ML}_X\right\| > 2\epsilon_p$}{
				$S$ : $\mathbb{F}_S \leftarrow X$, $\mathbb{V}_S = \mathbb{V}_S \setminus X$\;
			}
		}
  }
}
\caption{Multilateration (ML) test}
\label{alg:vt_ml}
\end{algorithm}
\decmargin{2pt}

\begin{table}[t]
\begin{center}
\renewcommand{\arraystretch}{1.3}
\caption{Summary of security analysis in a generic network topology}
\label{tab:security_1a}
\begin{tabular}{| p{0.3\columnwidth} | l | p{0.14\columnwidth} | p{0.17\columnwidth} | l |}
\hline
\backslashbox{$X$}{~~~~~~~~~$\left|\mathbbm{N}_S\setminus X\right|$} & 0 & 1 & 2 & 3+ \\
\hline
Correct &
$\mathbbm{U}_S$ &
$\mathbbm{U}_S$ &
$\mathbbm{V}_S$ &
$\mathbbm{V}_S$ \\
\hline
Unknowledgeable adversary &
$\mathbbm{F}_S$ &
$\mathbbm{F}_S$ &
$\mathbbm{F}_S$ &
$\mathbbm{F}_S$ \\
\hline
Knowledgeable adversary &
$\mathbbm{U}_S$ &
$\mathbbm{U}_S$ (0.5) $\mathbbm{F}_S$ (0.5) &
$\mathbbm{U}_S$ (0.165) $\mathbbm{F}_S$ (0.835) &
$\mathbbm{F}_S$ \\
\hline
\end{tabular}
\end{center}
\end{table}

\begin{figure}[ht]
\centering
\includegraphics[width=0.75\columnwidth]{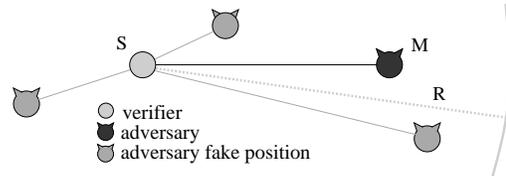}
\caption{
If $M$ knows $S$'s position,  it can advertise any
fake position, provided its distance from $S$ is
at most equal to $R$.}
\label{fig:security_1a0n}
\end{figure}

\begin{figure}[ht]
\centering
\includegraphics[width=0.65\columnwidth]{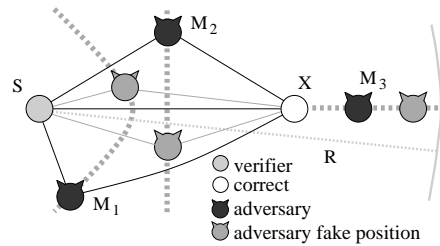}
\caption{
$M_1$, $M_2$, and $M_3$ depict different situations in
which a single adversary
can be. In the general case (as $M_1$),
a knowledgeable adversary that correctly guessed the verifier's identity
can pass all tests if its fake position
is on a hyperbola with foci in $S$, $X$, passing by $M_1$.
Particular cases that determine a degeneration of the hyperbola are:
(i) the adversary is equidistant
from $S$ and $X$ (as $M_2$), constraining the fake position
on the symmetry axis of $S$ and $X$;
(ii) the adversary is aligned with $S$ and $X$ (as $M_3$), and not between them:
then, the fake location needs to be on the same line, between $X$ and
a point at distance $R$ from $S$.}
\label{fig:security_1a1n}
\end{figure}

\begin{figure}[ht]
\centering
\includegraphics[width=0.6\columnwidth]{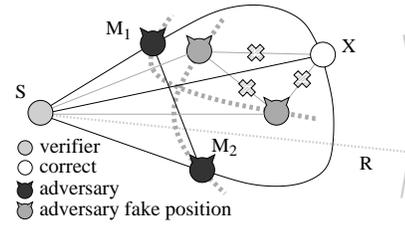}
\caption{
Clique of four nodes: the verifier $S$, a correct neighbor $X$,
and two adversaries ($M_1$, $M_2$). $M_1$ ($M_2$) announces a fake position
along a hyperbola with foci on $p_S$ and $p'_{M_2}$ ($p'_{M_1}$).
However, the latter information is
fake, leading to  a mismatch in the cross-check on  ($M_1$,$M_2$).
Also, since each attacker can ``move'' at most one link other than that
with $S$, the checks on ($X$,$M_1$) and ($X$,$M_2$) fail as well.
Thus, $M_1$ and $M_2$ damage each other and are tagged as faulty.
$X$, although correct, is added to $\mathbb{F}_S$, since all neighbors it
shares with $S$ happen to be adversaries.}
\label{fig:security_ma_independent}
\end{figure}

\begin{figure}[ht]
\subfigure[Actual positions and links]{\label{fig:security_ma_colluding_1}
\includegraphics[width=0.43\columnwidth]{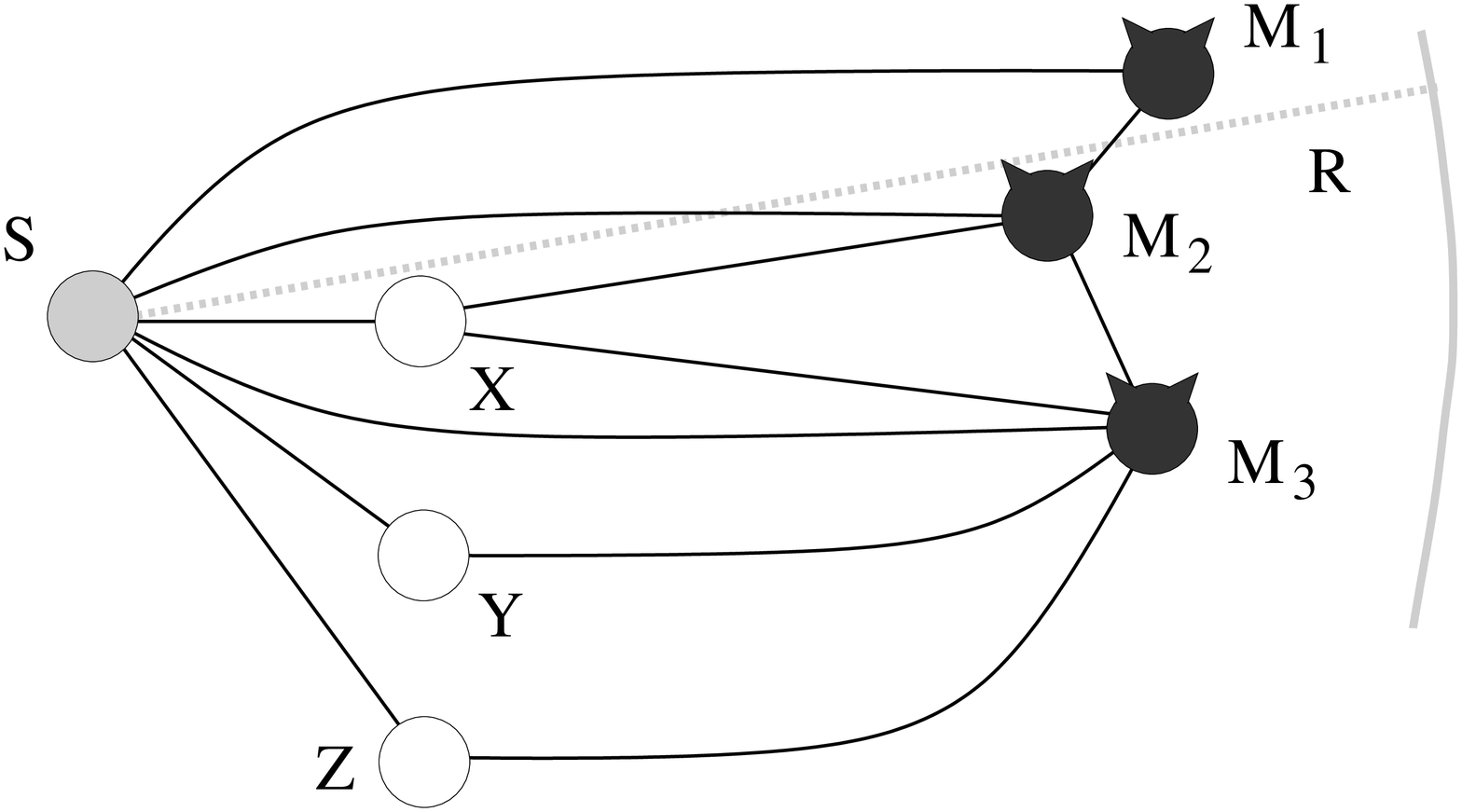}
}
\subfigure[Coordinated attack]{\label{fig:security_ma_colluding_2}
\includegraphics[width=0.43\columnwidth]{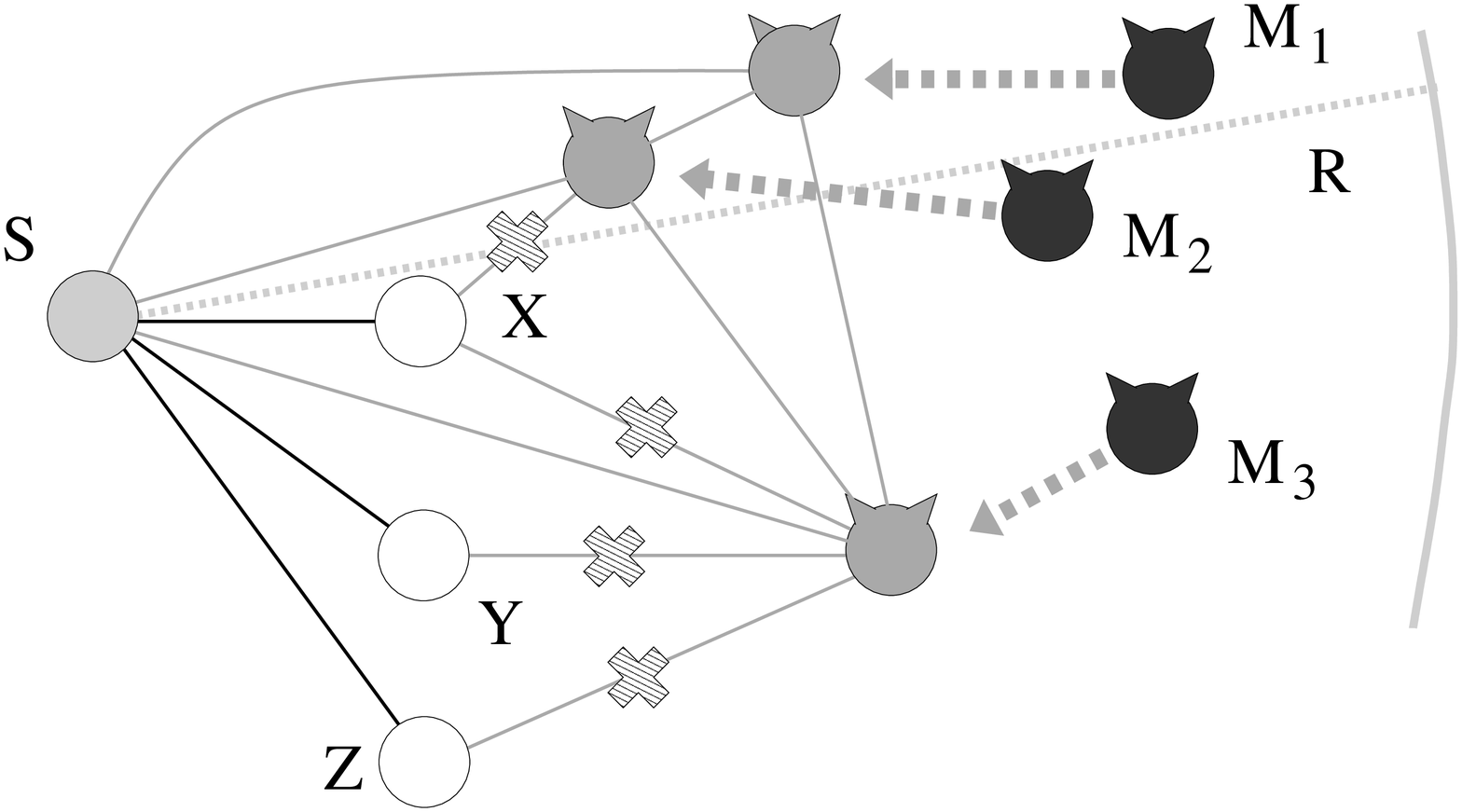}
}
\caption{
Coordinated attack by $M_1$, $M_2$, and $M_3$ against
$S$.
All links between adversaries appear consistent with the
false positions they advertise, but links with correct neighbors
$X$, $Y$, and $Z$ result in mismatches in the {\bf CS} test. $M_1$, sharing with $S$
two colluders but no correct nodes, results
as verified. The same holds for $M_2$, sharing with $S$ two colluders
and one correct node. $M_3$ is instead marked as faulty, thanks
to the three correct common neighbors.}
\label{fig:security_ma_colluding}
\end{figure}

\begin{figure*}[tb]
\subfigure[ ]{\label{fig:zurich_malw}
	\includegraphics[width=0.5\textwidth]{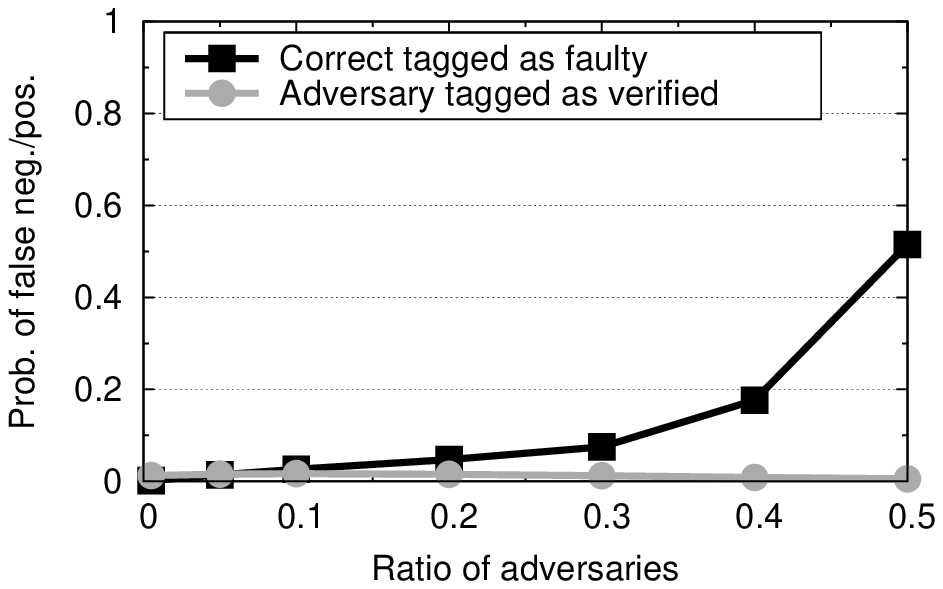}
}
\hspace{-0.5cm}
\subfigure[ ]{\label{fig:zurich_malu}

\includegraphics[width=0.5\textwidth]{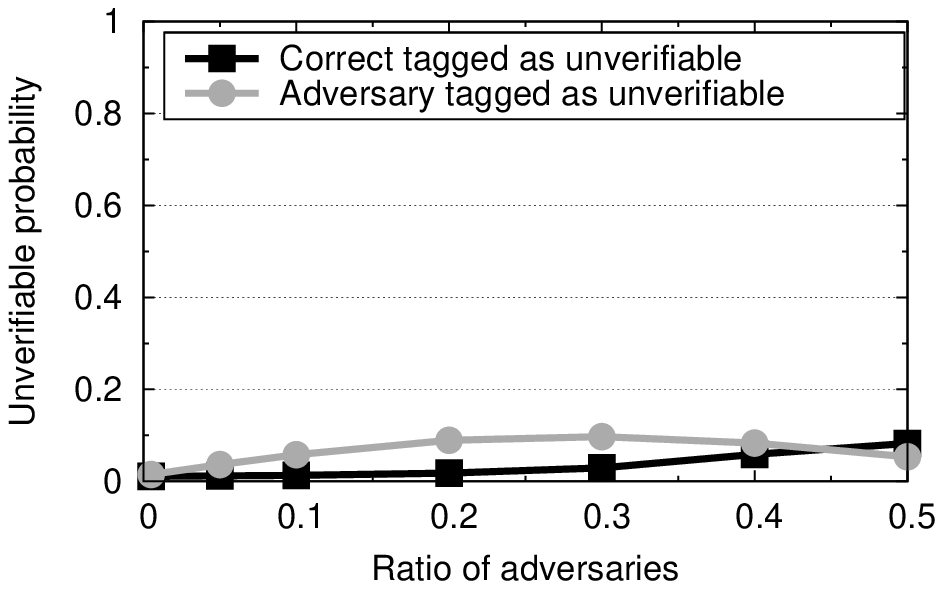}
}
\hspace{-0.5cm}
\subfigure[ ]{\label{fig:zurich_Rw}

\includegraphics[width=0.5\textwidth]{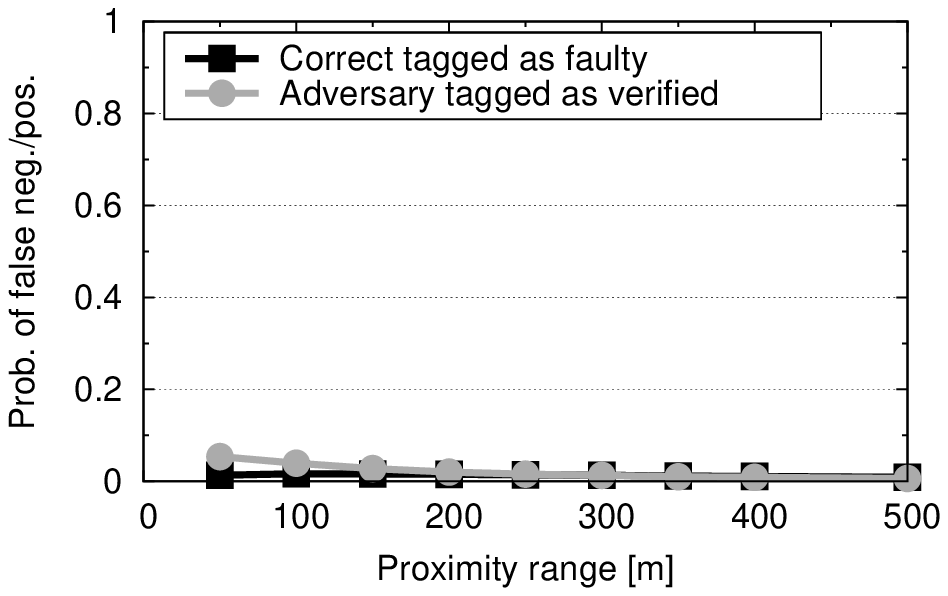}
}
\hspace{-0.5cm}
\subfigure[ ]{\label{fig:zurich_Ru}

\includegraphics[width=0.5\textwidth]{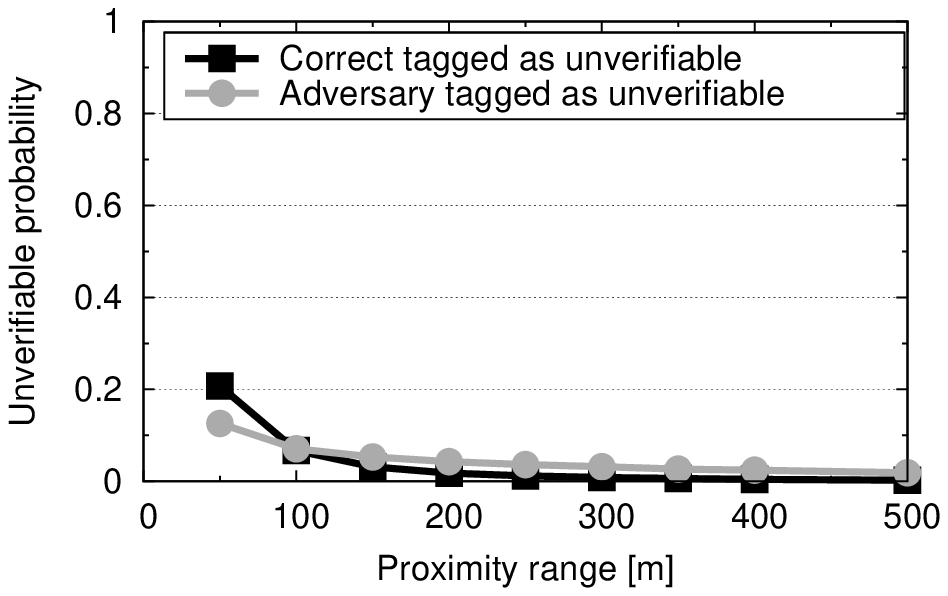}
}
\caption{Independent adversaries: probability of false negatives/positives and
probability of classifying a neighbor
as unverifiable. In (a) and (b), $R=250$~m while the ratio of adversaries varies;
in (c) and (d), the ratio of adversaries is 0.05 and the proximity range $R$ varies.}
\label{fig:zurich}
\end{figure*}

\begin{figure*}[tb]
\subfigure[ ]{\label{fig:coll_zurich_w}
	\includegraphics[width=0.5\textwidth]{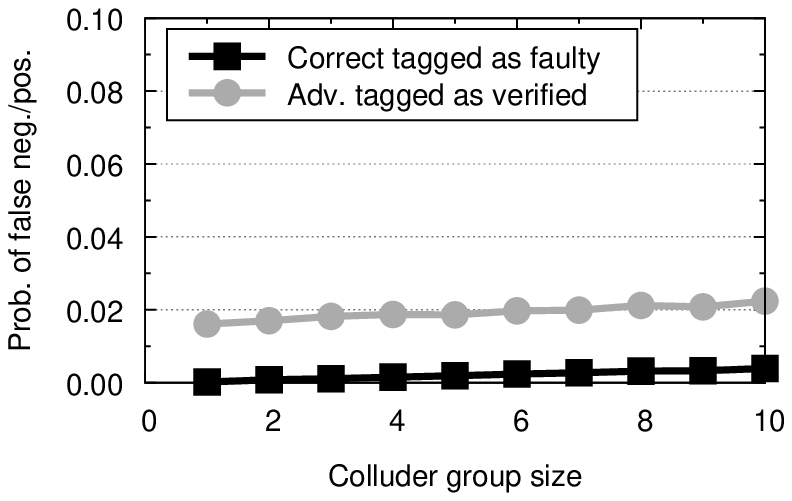}
}
\hspace{-0.5cm}
\subfigure[ ]{\label{fig:coll_zurich_u}

\includegraphics[width=0.5\textwidth]{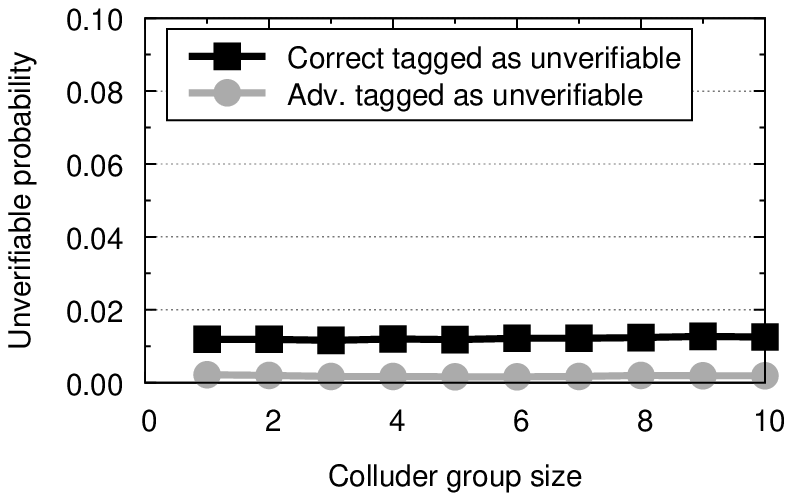}
}
\caption{Colluding adversaries: probability of false negatives/positives and probability of classifying a neighbor
as unverifiable,  for ratio of adversaries equal to 0.05, $R=250$~m, and varying group
size $\sigma$. }
\label{fig:coll}
\end{figure*}

\end{document}